# Effects of time aggregation, product aggregation, and seasonality in measuring bullwhip ratio

Hau Mike Ma [a], Jiazhen Huo [b,c,*], Yongrui Duan [b,c]

[a] *School of Mechanical Engineering, Tongji University, Shanghai, 200092, China*
[b] *School of Economics and Management, Tongji University, Shanghai, 200092, China*
[c] *Bosch-Tongji University Chair of Global Supply Chain Management, Shanghai, 200092, China*

**Abstract.**

The bullwhip study has received a lot of attention in the literature, but with conflicting results, especially in the context of data aggregation. In this paper, we investigate three widely studied factors in bullwhip measurement: time aggregation, product aggregation, and seasonality. In time aggregation, we decompose the variance into two components: the expectation of the subset variances and the variance of subset expectations, thus decomposing the bullwhip ratio into four components to explore the underlying mechanism of time aggregation. In product aggregation, the bullwhip ratio is analyzed in the context of products with either uncorrelated or correlated demands and orders. Seasonality is also examined to study its effect on the bullwhip ratio. Our key findings are: (a) Time aggregation can increase, decrease, or maintain the bullwhip ratio in different scenarios. (b) Aggregated bullwhip ratio of uncorrelated products is a weighted average of bullwhip ratios from individual products, with corresponding demand variance as the weights. However, aggregated bullwhip ratio of correlated products could break the boundaries. (c) Seasonality can be considered as a standalone product with a bullwhip ratio of one, which can drive the overall bullwhip ratio closer to one.

**Key words:** bullwhip ratio measurement, time aggregation, product aggregation, seasonality

* Corresponding author. School of Economics and Management, Tongji University, Shanghai, 200092, China.
E-mail address: huojiazhen519@163.com (J. Huo).

If you have any questions or comments, please send them to me at:
**ma_mike@tongji.edu.cn**

## 1. Introduction

Effect of aggregation on the bullwhip effect has been meritoriously studied in the literature. Lee et al. (1997) modeled cases of order batching to demonstrate its contribution to the bullwhip effect. Cachon et al. (2007) mentioned that aggregation influence on the bullwhip effect depends on the correlation of production and demand across the units being aggregated. Chen and Lee (2012) developed a simple set of formulas to describe the traditional bullwhip measure as a combined influence of several drivers such as seasonality, time aggregation, product/location aggregation. Yao et al. (2020) studied the aggregation influence on bullwhip effect empirically using a dataset from a large supermarket chain, reporting that aggregated bullwhip effect ratios underestimate the bullwhip effect. However, the effect of aggregation on the bullwhip is still inconsistent.

There are two primary approaches for measuring the bullwhip phenomenon in academic studies. The first approach captures the distortion in information flow upstream by comparing the variance of orders with the variance of demand (Lee et al., 1997), and the second approach compares the variance of order receipts or shipments with the variance of sales. Therefore, the bullwhip effect can be represented as either the ratio or the difference between two variances. Chen and Lee have mentioned that the bullwhip ratio is a more effective metric than the bullwhip difference when analyzing different products (Ha and Tang, 2017). In this paper, we will focus on the measurement of the bullwhip ratio.

Our paper starts with the essence of the bullwhip ratio to study the conflicting results of aggregation effect on the bullwhip ratio. We first explore the underlying factors of the bullwhip ratio based on the law of total variance, and then study effect of time aggregation on the bullwhip ratio. Second, we consider products with either uncorrelated or correlated orders and demands to study the effect of product aggregation on the bullwhip ratio. Finally, we examine the influence of seasonality on the bullwhip ratio. The purpose of our paper is not to question the existence of the bullwhip effect, but to examine the aggregation effect on the bullwhip ratio, i.e., whether time aggregation, product aggregation, and seasonality increase, decrease, or maintain the non-aggregated bullwhip ratio. The major objectives of our work are threefold: (i) decompose the bullwhip ratio using the law of total variance to delve into time aggregation; (ii) study products with or without correlation to gain insight into influence of product aggregation on the bullwhip ratio; and (iii) examine seasonality to explore its effect on the bullwhip ratio.

The remainder of this paper is structured as follows. In Section 2, we review the literature on bullwhip ratio modeling and aggregation effect estimation. In Section 3, we use the law of total variance to decompose non-aggregated bullwhip ratio to investigate the effect of time aggregation on the bullwhip ratio. In Section 4, effect of product aggregation on the bullwhip ratio is studied. In Section 5, seasonality is considered to study its effect on the bullwhip measurement. In Section 6, numerical analysis is conducted. Finally, Section 7 presents our concluding remarks.

If you have any questions or comments, please send them to me at:
**ma_mike@tongji.edu.cn**

## 2. Related literature review

### 2.1 Modeling the bullwhip ratio

There are several normative approaches to modeling demand processes for the bullwhip ratio. Lee et al. (1997) considered an autoregressive AR(1) demand process for modeling the bullwhip effect. Graves (1999) modeled demand using an integrated moving-average IMA (0,1,1) process. Dejonckheere et al. (2003) modeled a supply chain as an engineering system and studied the bullwhip effect under certain system replenishment rules. Chen and Lee (2009) proposed a demand model that generalizes the demand processes which evolves according to the martingale model of forecast evolution (MMFE) process.

These studies of the bullwhip phenomenon typically model demands first, then derive orders, and finally calculate the bullwhip ratio. In this paper, we focus on examining the aggregation effect on the bullwhip ratio, where both the aggregated and non-aggregated bullwhip ratios are derived from the same demand process. As a result, our analysis emphasizes direct variance calculations rather than delving into the detailed characteristics of the demand process. As defined by Blitzstein and Hwang (2019), variance serves as a single-number summary of the distribution of a random variable. Using the law of total variance, we decompose the variances of demands or orders into two components. Subsequently, we further decompose the bullwhip ratio into four distinct components. This approach enables us to construct both the aggregated and non-aggregated bullwhip ratios using the same formula, thereby facilitating the analysis of time aggregation effects.

### 2.2 Aggregation effect on the bullwhip ratio

In this section, we review the literature of aggregation effect on the bullwhip ratio with three subtopics: time aggregation, product aggregation, and seasonality.

Lee et al. (1997) studied the effect of batching on the bullwhip ratio considering three ordering scenarios. Chen and Lee (2012) proposed a general theoretical framework to explain various empirical observations, showing that time aggregation can mask the bullwhip effect, supported by a weekly, SKU level dataset from a European retail store during a one-year period. Cachon et al. (2007) analyzed a U.S. industry-level dataset and find that seasonality attenuates the bullwhip effect and report that aggregation may not have to cause bias. Bray and Mendelson (2012) examined the bullwhip effect using a sample of 4689 public firms from 1974 to 2008 and find that demand signals from lead times contribute to the bullwhip effect. Shan et al. (2013) investigated the bullwhip effect using quarterly data incorporating over 1200 public firms in China from 2002 to 2009, collecting interim and annual financial data and calculating the quarterly, half-yearly, and yearly bullwhip ratios, respectively to prove that bullwhip ratios does not change significantly under time aggregation. Duan et al. (2015) found that under substitute products, the bullwhip effect is not only affected by a product's own factors but also by those of its substitute products. Jin et al. (2016) studied intra-industry bullwhips, reporting that time aggregation tends to reduce the bullwhip. Yao et al. (2020) reported that aggregated bullwhip effect ratios by store and by time are lower than non-aggregated bullwhip ratios. Lee et al. (2023) used an annual dataset obtained from the Export–Import Bank of Korea and reported that subsidiary-level data exhibit less product aggregation but more time aggregation.

Although previous studies have explored the impact of aggregation on the bullwhip ratio, their findings exhibit inconsistent results (refer to Table 1). Our paper is seeking to address the arbitrariness by providing a more rigorous analysis. First, we employ a theoretical approach to dissect the fundamental nature of the bullwhip ratio and identify the underlying factors of time aggregation. Second, we examine uncorrelated and correlated products to explore the effect of product aggregation on the bullwhip ratio. Third, we also examine the effect of seasonality on the bullwhip phenomenon and find similar results to Chen and Lee (2012).

**Table 1. Summary of key literature on time aggregation, product aggregation, and seasonality**

| Paper | Method | Data source | Time aggregation | Product aggregation | Seasonality | Findings |
|---|---|---|---|---|---|---|
| Cachon et al. (2007) | Empirical | Monthly, industry level data in the United States during Jan 1992 to Feb 2006 | √ (0) |  | √ (-) | Seasonality attenuates the bullwhip effect and claim the aggregation may not have to cause bias. |
| Chen and Lee (2012) | Analytical | A data set from Rob Broekmeulen of Eindhoven University of Technology, which contains weekly sales and delivery data for six consumer products from a European retail store during a one-year period. | √ (-) | √ (?) | √ (-) | Seasonality dampens bullwhip effect; Time aggregation masking bullwhip effect; Product/location aggregation inconclusive. |
| Bray and Mendelson (2012) | Empirical | Quarterly data of 4,689 U.S. public firms during 1974–2008 |  |  | √ (-) | Predict a negative seasonal bullwhip. |
| Shan et al. (2013) | Empirical | Quarterly data of 1,200 public firms in China from 2002 Q1 to 2009 Q2 | √ (0) |  |  | Bullwhip ratios does not change significantly under time aggregation. |
| Duan, Yao, and Huo (2015) | Empirical | Daily data of a large Chinese supermarket chain from April to October 2011 |  | √ (-) |  | Bullwhip effect is affected by substitute product's price changes. |

| Jin et al. (2016) | Analytical | | √ (+/-) | | Dampen the bullwhip if the ratio is greater than one at the shorter time aggregation level, but a new finding is that it amplifies the bullwhip if the ratio is less than one. |
| Yao, Duan, and Huo (2020) | Empirical | Daily data of a large Chinese supermarket chain from April to October 2011 | √ (-) | √ (-) | Aggregated bullwhip effect ratios by store and by time are lower than non-aggregated bullwhip effect ratios |
| Lee et al. (2023) | Empirical | Annual dataset obtained from the Export–Import Bank of Korea between 2006 and 2013. | √ (+) | √ (-) | Subsidiary-level data exhibit less product aggregation but more time aggregation. |

Note: "√" means that this type of aggregation was studied in the paper. "0" means that aggregation has no obvious effect on the bullwhip ratio, "+" means that aggregation could increase the non-aggregated bullwhip ratio, and "-" means aggregation could decrease the non-aggregated bullwhip ratio. "?" means that effect of aggregation is inconclusive. In the literature, there are various expressions of aggregation effect. In this paper, we classify the aggregation effects into three types: **"increase"** (referred to in the literature as amplify, be higher than, more, etc.), **"decrease"** (referred to as dampen, reduce, attenuate, be lower than, less, etc.), and **"maintain"** (referred to as having no effect, no significant effect, etc.).

## 3. Time aggregation analysis

### 3.1 Decomposing the bullwhip ratio

In this section, we examine the effect of time aggregation on the bullwhip ratio, using realized demand and order sequences. We consider a supply chain member managing customer demands and placing orders to suppliers for a single product, under the assumptions of unlimited capacity and no batch-ordering.

Suppose we have observed a demand sequence of size $T$ from random variable $D_t$, denoted as $\{d_t\}_{t=1}^T = \{d_1, d_2, \dots, d_T\}$. Divide this sequence into $M$ subsets, with each subset containing $k$ data points such that $T = M \cdot k$. Without losing generality, assume that $M$ and $k$ are integers. Let each subset be represented as $\left\{d_t^{(j)}\right\}_{t=1}^k$ for $j = 1, 2, \dots, M$. The mean of the $j$-th subset is denoted as $\bar{d}^{(j)} = \frac{1}{k}\sum_{t=1}^k d_t^{(j)}$ and the variance of the $j$-th subset is denoted as $\sigma_{d_t^{(j)}}^2 = \frac{1}{k}\sum_{t=1}^k \left(d_t^{(j)} - \bar{d}^{(j)}\right)^2$. Define $\sigma_{d_{within}}^2$ as the expected value of all subset variances, and $\sigma_{d_{within}}^2 = \frac{1}{M}\sum_{j=1}^M \sigma_{d_t^{(j)}}^2$. Define $\sigma_{\bar{d}^{(j)}}^2$ as the variance of all the subset means. The variance of non-aggregated demands is denoted as $\sigma_{d_{total}}^2$.

Using the law of total variance in Blitzstein and Hwang (2019, p.434), we can express the total variance as the sum of expectation of subset variances and the variance of subset means.

Thus,

$$\sigma_{d_{total}}^2 = \sigma_{d_{within}}^2 + \sigma_{\bar{d}^{(j)}}^2 \tag{1}$$

Similarly, we can have the variance for the order sequence:

$$\sigma^2_{o_{total}} = \sigma^2_{o_{within}} + \sigma^2_{\bar{o}(j)} \tag{2}$$

**Proposition 1**. *In the context described above, variance of orders, variance of demands, and the bullwhip ratio can be decomposed into:*

$$\frac{\sigma^2_{o_{total}}}{\sigma^2_{d_{total}}} = \frac{\sigma^2_{o_{within}} + \sigma^2_{\bar{o}(j)}}{\sigma^2_{d_{within}} + \sigma^2_{\bar{d}(j)}} \tag{3}$$

Formula (3) is a general formula of the bullwhip ratio, which gives a high-level holistic view on bullwhip ratio components by decomposing the total variance into expectation of subset variances and variance of subset means. As illustrated in Figure 1, $\sigma_{o_{total}}$ is the hypotenuse, $\sigma_{o_{within}}$ and $\sigma_{\bar{o}(j)}$ are the right-angle sides; Similarly, $\sigma_{d_{total}}$ is the hypotenuse, $\sigma_{d_{within}}$ and $\sigma_{\bar{d}(j)}$ are the right-angle sides. Consequently, the bullwhip ratio is the ratio of these square areas. As depicted in Figure 1, the total bullwhip ratio $R_1 = \frac{\sigma^2_{o_{total}}}{\sigma^2_{d_{total}}}$ can be expressed by $R_3 = \frac{\sigma^2_{o_{within}}}{\sigma^2_{d_{within}}}$ and $R_2 = \frac{\sigma^2_{\bar{o}(j)}}{\sigma^2_{\bar{d}(j)}}$. Interestingly, $R_2$ is the aggregated bullwhip ratio under time aggregation and a detailed proof will be given in Section 3.3.

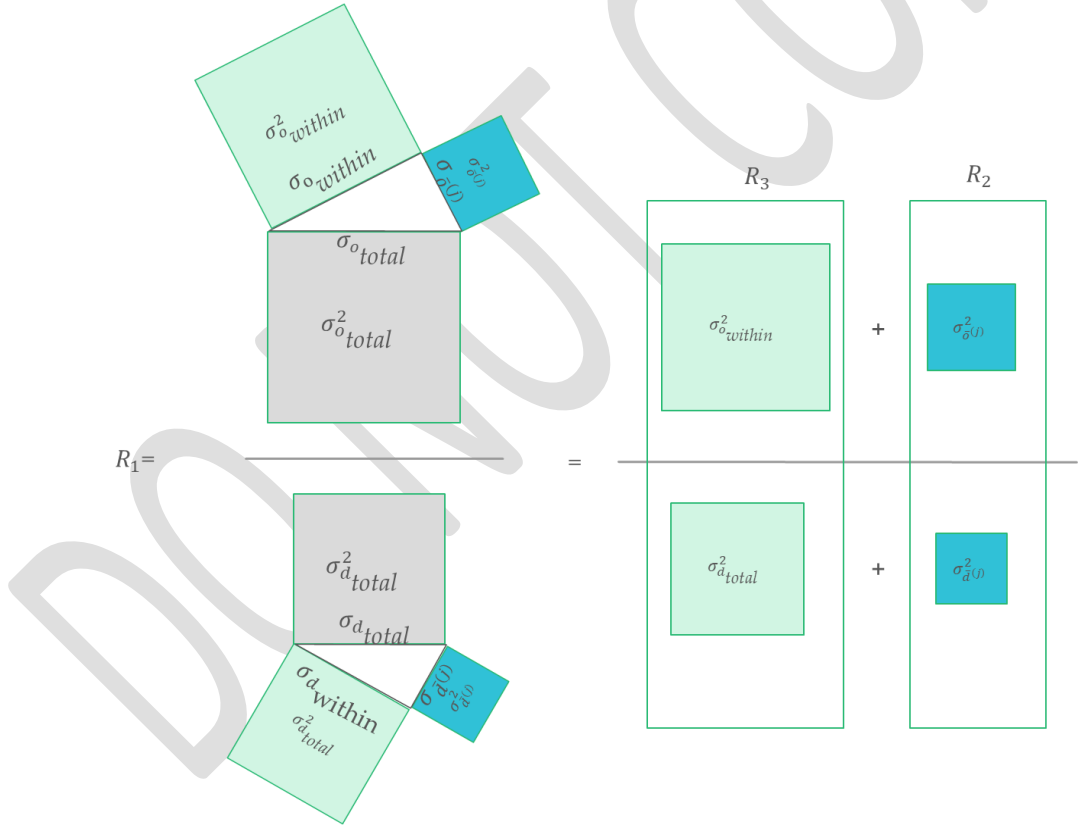

**Figure 1. Illustration of bullwhip ratio decomposition based on the law of total variance**

In time aggregation, the bullwhip ratio can be decomposed into granular components to provide insights into supply chain performance. High within-subset variability suggests that there is significant fluctuation in orders or demands at the granular level, which could indicate inefficiencies or instability within subsets. Between-subset variability shows how much the order means or demand means differ across different subsets. Supply chain managers can take more targeted interventions.

For example, if within-subset variability is the primary driver, efforts might focus on process improvements or stabilizing specific segments. If between-subset variability is more significant, strategic decisions might involve revisiting the overall supply chain strategy or demand planning approaches. Formula (3) could re-evaluate the bullwhip ratio as a performance metric. It shows that a bullwhip ratio close to one does not necessarily mean optimal performance. It highlights that the bullwhip ratio can be manipulated by time aggregation, and a low bullwhip ratio may still mask underlying inefficiencies. Thus, the bullwhip ratio should be interpreted in context, considering both within-subset and between-subset variances.

### 3.2 Breaking down $\sigma^2_{d_{within}}$ and $\sigma^2_{o_{within}}$ in Formula (3)

In this part, we will delve into the expectation of subset variances $\sigma^2_{d_{within}}$ and $\sigma^2_{o_{within}}$ to gain a deeper understanding of Formula (3) and its implications for the bullwhip ratio. The expectation of subset variances $\sigma^2_{d_{within}}$ represents the average within-subset variability of demands, which is $\frac{1}{M}\sum_{j=1}^{M}\sigma^2_{d_t^{(j)}}$. Similarly, $\sigma^2_{o_{within}}$ captures the average within-subset variability for orders, which is $\frac{1}{M}\sum_{j=1}^{M}\sigma^2_{o_t^{(j)}}$.

**Proposition 2.** *Assume that order sequence and demand sequence are divided into M subsets (without losing generality, assuming each subset with the same amount of elements), the bullwhip ratio of the jth subset is denoted as $r_j$, and $r_j = \frac{\sigma^2_{o_t^{(j)}}}{\sigma^2_{d_t^{(j)}}}$ ; the weighted average of $r_j$ is denoted as $w_j$, and $w_j = \frac{\sigma^2_{d_t^{(j)}}}{\sum_{j=1}^{M}\sigma^2_{d_t^{(j)}}}$, the ratio of average subset order variance and demand variance*

$$\frac{\sigma^2_{o_{within}}}{\sigma^2_{d_{within}}} = r_1 w_1 + \cdots + r_M w_M \qquad (4)$$

Formula (4) gives a clear demonstration of how each subset contributes to the overall bullwhip effect. Subsets with higher individual bullwhip ratios $r_i$ will have a more significant impact on the total bullwhip ratio if they are assigned larger weights. Each subset's contribution to the total bullwhip ratio is weighted by the variance of demand within that subset. This means that subsets with higher demand variance will have a greater influence on the overall bullwhip ratio. This is critical because it underscores that not all subsets are equally impactful.

### 3.3 Deriving the aggregated bullwhip ratio based on Formula (3)

We denote the aggregated demand of the $j$th subset of the demand sequence as $d_{\text{agg}}^{(j)}$, and $d_{\text{agg}}^{(j)} = \sum_{t=1}^{k} d_t^{(j)}$. Let $\bar{d}^{(j)} = \frac{1}{k}\sum_{t=1}^{k} d_t^{(j)}$ be the mean of the $j$-th subset of demands. We can get $d_{\text{agg}}^{(j)} = k\bar{d}^{(j)}$.

Similarly, we denote the aggregated order of the *j*th subset of the order sequence as $o_{agg}^{(j)}$, and $o_{agg}^{(j)} = \sum_{t=1}^{k} o_t^{(j)}$. Let $\bar{o}^{(j)} = \frac{1}{k}\sum_{t=1}^{k} o_t^{(j)}$ be the mean of the *j*-th subset of orders. We can get $o_{agg}^{(j)} = k\bar{o}^{(j)}$.

**Proposition 3**. *Let $\sigma^2_{d_{agg}^{(j)}}$ denote the variance of the aggregated demands across M subsets, $\sigma^2_{o_{agg}^{(j)}}$ denote the variance of the aggregated orders across M subsets. Let $\sigma^2_{\bar{d}^{(j)}}$ denote the variance of the averaged demands across M subsets, and $\sigma^2_{\bar{d}^{(j)}}$ denote the variance of averaged orders across M subset, the aggregated bullwhip ratio*

$$\frac{\sigma^2_{o_{agg}^{(j)}}}{\sigma^2_{d_{agg}^{(j)}}} = \frac{\sigma^2_{\bar{o}^{(j)}}}{\sigma^2_{\bar{d}^{(j)}}} \tag{5}$$

Interestingly, the aggregated bullwhip ratio $\frac{\sigma^2_{o_{agg}^{(j)}}}{\sigma^2_{d_{agg}^{(j)}}}$ which equals to $\frac{\sigma^2_{\bar{o}^{(j)}}}{\sigma^2_{\bar{d}^{(j)}}}$, can be found in the non-aggregated bullwhip ratio in formula (3). Thus, we can compare the aggregated bullwhip ratio and the non-aggregated bullwhip ratio in Section 3.4.

The aggregated bullwhip ratio $\frac{\sigma^2_{o_{agg}^{(j)}}}{\sigma^2_{d_{agg}^{(j)}}}$ is the same to $\frac{\sigma^2_{\bar{o}^{(j)}}}{\sigma^2_{\bar{d}^{(j)}}}$, demonstrating a key insight: the aggregated bullwhip ratio can be expressed within the non-aggregated bullwhip ratio.

**3.4 Compare time-aggregated bullwhip ratio and non-aggregated bullwhip ratio**

In this section, we will target the effect of time aggregation on the bullwhip measurement. The setting is the same to those in Sections 3.1 through 3.3. Now denote $r_{non-agg} = \frac{\sigma^2_{o_{total}}}{\sigma^2_{d_{total}}}$, $r_{avg} = \frac{\sigma^2_{\bar{o}^{(j)}}}{\sigma^2_{\bar{d}^{(j)}}}$, $r_{agg} = \frac{\sigma^2_{o_{agg}^{(j)}}}{\sigma^2_{d_{agg}^{(j)}}}$, and $r_{within} = \frac{\sigma^2_{o_{within}}}{\sigma^2_{d_{within}}}$, we can get $r_{avg} = r_{agg}$ based on Formula (5).

To answer the puzzling question whether time aggregation will increase, decrease or maintain the non-aggregated bullwhip ratio. We need to compare no-aggregated bullwhip ratio $r_{non-agg}$ and the aggregated bullwhip ratio $r_{agg}$.

**Proposition 4.** *Given the context above*

$$If\ r_{avg} > r_{within}, r_{agg} > r_{non-agg} \tag{6}$$

$$If\ r_{avg} < r_{within}, r_{agg} < r_{non-agg} \tag{7}$$

$$If\ r_{avg} = r_{within}, r_{agg} = r_{non-agg} \tag{8}$$

The relationship between non-aggregated bullwhip ratio $r_{non-agg}$ and the aggregated bullwhip ratio $r_{agg}$ is decided by the ratio of subset mean variances $r_{avg}$ and the within-subset bullwhip ratio $r_{within}$. Formula (6) shows that when the ratio of subset mean variances is larger than the within-subset bullwhip ratio, time aggregation will increase the bullwhip ratio; Formula (7) shows that when the ratio of subset mean variances is smaller than the within-subset bullwhip ratio, time aggregation will decrease the bullwhip ratio; Formula (8) shows that when the subset mean variances is equal to the within-subset bullwhip ratio, time aggregation will maintain the bullwhip ratio.

Proposition 4 provides a theoretical basis for understanding results reported in the literature. For instance, Table 2 is the reported results from Yao et al. (2020), showing that the bullwhip ratio does not decrease monotonically under time aggregation, outliers are highlighted in red. Similarly, Chen and Lee (2012) analyzed weekly sales and delivery data for six consumer products from a European retail store (in Table 3). They observed that the bullwhip ratio generally decreases as the aggregation period increases. However, outliers such as applesauce and tea bags were noted. Conflicting results in the literature can be reasonably explained using formulas (6), (7), and (8). Lee et al. (2023) reported that subsidiary-level data exhibit more pronounced effects of time aggregation, which can be understood through the lens of formula (6). Chen and Lee (2012) and Yao et al. (2020) noted that time aggregation can mask or reduce the bullwhip effect, as explained by formula (7). Cachon et al. (2007) and Shan et al. (2013) observed that the bullwhip ratio does not change significantly under time aggregation, which corresponds to the scenario described by formula (8). By decomposing the total bullwhip ratio into within-subset and between-subset components, we offer a normative approach to understand how time aggregation influences the bullwhip ratio. This decomposition provides a clearer explanation for various scenarios observed in empirical studies, offering a more nuanced understanding of how time aggregation can either increase or decrease the bullwhip effect in different contexts.

**Table 2. Estimation of bullwhip effect by product category from Yao et. al (2020)**

| Estimations of bullwhip effect by product category (BW-HYD-FD) | | | | |
|---|---|---|---|---|
| **Product category** | **No aggregation (daily data)** | **One-week aggregation** | **Two-week aggregation** | **Four-week aggregation** |
| Juices | 48.79 | 26.49 | **27.73** | **28.89** |
| Potato chips | 77.42 | 19.25 | 15.71 | 12.45 |
| Tea drinks | 19.54 | 13.67 | 9.76 | 7.24 |
| Kids' tooth pastes | 45.63 | 31.97 | 31.24 | 14.76 |
| Tooth pastes | 66.75 | 37.12 | 30.26 | 23.48 |
| Tooth brushes | 70.02 | 27.32 | 18.37 | 14.17 |
| Mouth wash | 21.41 | 15.99 | 15.60 | **17.22** |
| Facial tissues | 24.41 | 11.83 | 7.41 | 3.81 |
| Wipes | 203.53 | 131.85 | **155.86** | 131.91 |
| Toilet papers | 23.41 | 10.54 | 6.45 | 5.30 |
| Baby wipes | 9.52 | 7.18 | **8.16** | 6.68 |
| Shampoos | 20.02 | 14.05 | 9.50 | 6.00 |
| Detergents | 19.07 | 9.46 | **11.15** | 11.11 |
| Vinegar | 20.82 | 12.42 | 11.94 | **12.13** |
| Cooking oil | 20.54 | 13.95 | 9.94 | 5.16 |

Table 3. Product aggregation by product and time, Chen and Lee (2012)

| Bullwhip ratio at a European retail store | | | | |
|---|---|---|---|---|
| Product | Pack size | Weekly | Biweekly | Four-week |
| Applesauce | 12 | 1.60 | **1.76** | 1.17 |
| Mineral water | 6 | 1.87 | 1.51 | 1.32 |
| Peanut butter | 12 | 1.50 | 1.25 | 1.08 |
| Stroopwafels | 15 | 1.31 | 1.23 | 1.17 |
| Sugar | 10 | 3.04 | 2.68 | 2.03 |
| Tea bags | 15 | 2.09 | **2.46** | 1.70 |
| Aggregate | | 2.35 | **2.41** | 1.99 |

**4. Product aggregation analysis**

In Section 4.1 and Section 4.2, we will analyze the effect of product aggregation on the bullwhip ratio, with uncorrelated or correlated demands and orders respectively. Consider a supply chain member facing demands from customers and placing orders to suppliers, with no capacity limit and no batch-ordering. Consider $N$-product demand and order quantities as $D_t^N = \sum_{n=1}^{N} D_{t,n}$ and $O_t^N = \sum_{n=1}^{N} O_{t,n}$ (Chen and Lee, 2012), where $D_{t,n}$ and $O_{t,n}$ are variables of individual product demands and orders respectively.

**4.1 Product aggregation considering uncorrelated products**

Consider products are uncorrelated in demands and orders. The individual product demand and order are denoted as $D_{t,n}$ and $O_{t,n}$ respectively, and the aggregated demand and order are denoted as $D_t^N = \sum_{n=1}^{N} D_{t,n}$ and $O_t^N = \sum_{n=1}^{N} O_{t,n}$. Define individual product bullwhip ratio as $r_n$, and $r_n = \frac{\text{Var}(O_{t,n})}{\text{Var}(D_{t,n})}$. Denote row vector $\boldsymbol{r}$ as all individual bullwhip ratios, and $\boldsymbol{r} = [r_1\ r_2\ \dots\ r_n]$. Define the weight $w_n$ for each product $n$ as the proportion of its demand variance to the total demand variance, and $w_n = \frac{\text{Var}(D_{t,n})}{\sum_{n=1}^{N} \text{Var}(D_{t,n})}$. Define a column vector $\boldsymbol{w}$, and $\boldsymbol{w} = [w_1\ w_2\ \dots\ w_n]^T$.

**Proposition 5.** *Assume no capacity limit and no batch ordering. Given N products with uncorrelated demands and orders, the aggregated bullwhip ratio is the weighted average of individual bullwhip ratios.*

$$\frac{\text{var}(O_t^N)}{\text{var}(D_t^N)} = \frac{\sum_{n=1}^{N} \text{Var}(O_{t,n})}{\sum_{n=1}^{N} \text{Var}(D_{t,n})} = \boldsymbol{rw} \tag{9}$$

$$\min(r_1, r_2, \dots, r_n) \leq \frac{\text{var}(O_t^N)}{\text{var}(D_t^N)} \leq \max(r_1, r_2, \dots, r_n) \tag{10}$$

Where $r_n = \frac{\text{var}(O_{t,n})}{\text{var}(D_{t,n})}$, $\boldsymbol{r} = [r_1\ r_2\ \dots\ r_N]$, $w_n = \frac{Var(D_{t-1,n})}{\sum_{n=1}^{N} Var(D_{t,n})}$, and $\boldsymbol{w} = [w_1\ w_2\ \dots\ w_N]^T$.

Aggregated bullwhip ratio of uncorrelated products is a weighted average of the individual ratios $r_n$ where the weights are the variances of the demands. Therefore, the aggregated bullwhip ratio is more heavily influenced by products with larger demand variance and higher individual bullwhip ratios. Aggregated bullwhip ratio of multiple uncorrelated products is bounded by the minimum and maximum bullwhip ratios of the individual products. The principles of product aggregation can be extended from product-level aggregation into company-level and industry-level analyses. At the company level, the bullwhip ratio can be viewed as an aggregation of the ratios across different product lines. Similarly, at the industry level, the bullwhip ratio can be seen as an aggregation of the bullwhip ratios of the constituent companies.

### 4.2 Product aggregation considering correlated products

Consider products which can be correlated in demands and orders, with no capacity limit and no batch-ordering. Compared to Section 4.1, we leave out *t* in the notations without losing generality. Denote order of the *n*-th product as a random variable $O_n$, and the random vector $\boldsymbol{O}$ is composed of random variables of $O_n$ and $\boldsymbol{O} = [O_1, O_2, \ldots O_N]^T$. Similarly, denote demand of the *n*-th product as $D_n$ and the random vector $\boldsymbol{D}$ is composed of $N$ random variables of $D_n$ and $\boldsymbol{D} = [D_1, D_2, \ldots D_N]^T$. Denote the aggregated order of $N$ products as a random variable $O^N$, and $O^N = O_1 + O_2 + \cdots + O_N$. Similarly, denote the aggregated demand of $N$ products as a random variable $D^N$, and $D^N = D_1 + D_2 + \cdots + D_N$. The bullwhip ratio of each individual product is denoted as $r_n$ and $r_n = \frac{Var(O_n)}{Var(D_n)}$. The bullwhip ratio vector is denoted as $\boldsymbol{r}$, and $\boldsymbol{r} = [r_1, r_2, \ldots r_N]$. The bullwhip ratio for aggregated products is denoted as $r_{add}$, and $r_{add} = \frac{\text{Var}(O^N)}{\text{Var}(D^N)}$.

Variance-covariance matrix for $\boldsymbol{O}$:

$$\boldsymbol{\Sigma_O} = \begin{pmatrix} \text{Var}(O_1) & \text{Cov}(O_1, O_2) & \cdots & \text{Cov}(O_1, O_N) \\ \text{Cov}(O_2, O_1) & \text{Var}(O_2) & \cdots & \text{Cov}(O_2, O_N) \\ \vdots & \vdots & \ddots & \vdots \\ \text{Cov}(O_N, O_1) & \text{Cov}(O_N, O_2) & \cdots & \text{Var}(O_N) \end{pmatrix}$$

Variance-covariance matrix for $\boldsymbol{D}$:

$$\boldsymbol{\Sigma_D} = \begin{pmatrix} \text{Var}(D_1) & \text{Cov}(D_1, D_2) & \cdots & \text{Cov}(D_1, D_N) \\ \text{Cov}(D_2, D_1) & \text{Var}(D_2) & \cdots & \text{Cov}(D_2, D_N) \\ \vdots & \vdots & \ddots & \vdots \\ \text{Cov}(D_N, D_1) & \text{Cov}(D_N, D_2) & \cdots & \text{Var}(D_N) \end{pmatrix}$$

We denote $\lambda_{max}^O$ and $\lambda_{min}^O$ as the largest and smallest eigenvalues of $\boldsymbol{\Sigma_O}$ respectively, and $\lambda_{max}^D$ and $\lambda_{min}^D$ as the largest and smallest eigenvalues of $\boldsymbol{\Sigma_D}$ respectively.

**Proposition 6:** *Given products that can be correlated in orders or demands, we have*

$$\frac{\lambda_{min}^O}{\lambda_{max}^D} \leq r_{add} \leq \frac{\lambda_{max}^O}{\lambda_{min}^D} \tag{11}$$

It's important to note that $\lambda_{min}^O$ tends to be less than or equal to the smallest variance of $\text{Var}(O_n)$ and $\lambda_{max}^O$ tends to be greater than or equal to the largest variance $\text{Var}(O_n)$. Similarly, $\lambda_{min}^D$ tends to be less than or equal to the smallest variance of $\text{Var}(D_n)$ and $\lambda_{max}^D$ tends to be greater than or equal to the largest variance $\text{Var}(D_n)$. Therefore, $\frac{\lambda_{min}^O}{\lambda_{max}^D}$ could be less than the smallest individual bullwhip ratio $\min(r_n)$ and $\frac{\lambda_{max}^O}{\lambda_{min}^D}$ could be larger than the largest bullwhip ratio $\max(r_n)$. $r_{add}$ can be larger than the largest individual bullwhip ratio or smaller than the smallest individual bullwhip ratio.

Therefore, if products are correlated, the aggregated bullwhip ratio could bread the bounds derived in Section 4.1. Even through it is rarely observed in empirical studies, the results reported in Lee et al. (2023) can confirm the existence of this scenario.

## 5. Seasonality analysis

Seasonality is the predictable variability in demand throughout a year (Cachon et al., 2007). Chen and Lee (2012) showed that bullwhip ratio considering seasonality may go below one when there is a capacity limit in the system. In this section, we use the concept designed by Cleveland et al. (1990) to decompose a seasonal time series into two components: seasonal component and non-seasonal component, and seasonal and non-seasonal components are independent by nature.

Suppose we have observed a demand sequence of size $T$ from a random variable $D_t$, with $T$ observations as $\{d_t\}_{t=1}^T = \{d_1, d_2, \dots, d_T\}$. Based on Cleveland et al. (1990), $\{d_t\}_{t=1}^T$ can be decomposed into seasonality $\{s_t\}_{t=1}^T$ and seasonal adjusted demands $\{d'_t\}_{t=1}^T$. As we study the system property of seasonality, let random variable $S_{t,n}$ denote seasonality and random variable $D'_t$ denote seasonal adjusted demand. Similarly, we have an order sequence of size $T$ from random variable $O_t$, with $T$ observations as $\{O_t\}_{t=1}^T = \{O_1, O_2, \dots, O_T\}$. Let random variable $O'_t$ denote seasonal adjusted orders. As seasonality is the predictable variability in demand, thus order variability can be designed to be demand variability. Thus

$$\frac{Var(O_t)}{Var(D_t)} = \frac{Var(O'_t) + Var(S_t)}{Var(D'_t) + Var(S_t)}$$

**Proposition 7.** *Define $r_{adjusted}$ as $\frac{Var(O'_t)}{Var(D'_t)}$ and $r_{all}$ as $\frac{Var(O_t)}{Var(D_t)}$, given a deterministic seasonality,*

$$If\ r_{\text{adjusted}} > 1,\ r_{\text{adjusted}} > r_{all} > 1 \tag{12}$$

$$If\ r_{\text{adjusted}} < 1,\ r_{\text{adjusted}} < r_{all} < 1 \tag{13}$$

$$If\ r_{\text{adjusted}} = 1,\ r_{\text{adjusted}} = r_{all} = 1 \tag{14}$$

Seasonality can smoothen the bullwhip ratio towards one if bullwhip effect exists. This statement echoes to Proposition 2 of Chen and Lee (2012) that seasonality will lead to lower bullwhip effect. Involving seasonality could exacerbate the bullwhip ratio if the bullwhip ratio is less than one. This statement echoes to Proposition 3 of Chen and Lee (2012) that de-seasonalized bullwhip ratio could increase and reach one.

## 6. Numerical demonstration of time aggregation

In this section, we will numerically demonstrate key formulas derived in this paper.

### 6.1 Numerical demonstration of Formula (1)

In Table 4, we use a spreadsheet model to illustrate Formula (1). We simulate 6 demands: 9,5,8,6,7, and 9. The variance of these demands is 2.22. Interestingly, if $M = 6$, $k=1$, meaning we treat each order as its own subset, the variance within each subset is 0, and the subset means are simply the demands themselves. Consequently, the total variance remains 2.22. If $M = 3, k = 2$ where the 6 demands are divided into 3 subsets, the subset variances are as follows: the subset containing 9 and 5 has a variance of 4, the subset with 8 and 6 has a variance of 1, and the subset with 7 and 9 has a variance of 1. The expected value of these subset variances is 2. The subset means are 7,7, and 8, and the variance of these subset means is 0.22. Therefore, the total variance is still 2.22. Similarly, if $m = 2$, where the 6 demands are split into 2 subsets, the subset variances are 2.89 and 1.56, making the expected subset variance 2.22. The subset means in this case are both 7.33, resulting in a variance of the subset means of zero. Yet again, the total variance is 2.22. Finally, if $M = 1, k = 6$, treating all 6 demands as a single subset, the subset variance is 2.22, and since there is only one subset, the variance of the subset mean is 0. The total variance remains 2.22.

**Table 4. Numerical demonstration of variance decomposition formula with 6 demands**

| | Demonstration of variance of decomposition with 6 demands | | | | | | | | Var (D) |
|---|---|---|---|---|---|---|---|---|---|
| Order No. | | 1 | 2 | 3 | 4 | 5 | 6 | | |
| Volume | | 9 | 5 | 8 | 6 | 7 | 9 | | 2.22 |
| M=6, k=1 | Subset variance | - | - | - | - | - | - | Expectation of subset variances | - | |
| | Subset mean | 9.00 | 5.00 | 8.00 | 6.00 | 7.00 | 9.00 | Variance of the subset means | 2.22 | 2.22 |
| M=3, k=2 | Subset variance | | 4.00 | | 1.00 | | 1.00 | Expectation of subset variances | 2.00 | |
| | Subset mean | | 7.00 | | 7.00 | | 8.00 | Variance of the subset means | 0.22 | 2.22 |
| M=2, k=3 | Subset variance | | | 2.89 | | | 1.56 | Expectation of subset variances | 2.22 | |
| | Subset mean | | | 7.33 | | | 7.33 | Variance of the subset means | - | 2.22 |
| M=1, k=6 | Subset variance | | | | | | 2.22 | Expectation of subset variances | 2.22 | |
| | Subset mean | | | | | | 7.33 | Variance of the subset means | - | 2.22 |

## 6.2 Numerical demonstration of Formula (5.1)

In Table 5, we generate 6 orders randomly to show the key part of Formula (5.1), the relationship between variance of aggregated orders and variance of averaged orders. The orders are 9, 5, 8, 6,7, and 10. If we aggregate this order set into 3 subsets ($M = 3$), each with 2 numbers ($k = 2$), the aggregated subsets are 14.0, 14.0, and 17.0; the averaged subsets are 7.0, 7.0, and 8.5. Therefore, the variance of aggregated orders is 2, and the variance of averaged orders is 0.5, meaning the former is 4 times that of the latter. Similarly, if we aggregate this order set into 2 subsets ($M = 2$), each with 3 numbers ($k = 3$), variance of aggregated orders 0.250 is 9 times that of variance of averaged orders 0.028.

Table 5. Illustration of variance of aggregated orders and variance of averaged orders

| | Variance of subset aggregation vs. Variance of subset means | | | | | | | | |
|---|---|---|---|---|---|---|---|---|---|
| Order No. | | 1 | 2 | 3 | 4 | 5 | 6 | | Variance |
| Volume | Non-aggregated orders | 9 | 5 | 8 | 6 | 7 | 10 | | |
| M=3, k=2 | Aggregated orders, 2 data points in 1 subset | | 14.0 | | 14.0 | | 17.0 | Variance of aggregated orders | 2.000 |
| | Mean of each subset | | 7.0 | | 7.0 | | 8.5 | Variance of the subset means | 0.500 |
| M=2, k=3 | Aggregated orders, 3 data points in 1 subset | | | 22.0 | | | 23.0 | Variance of aggregated orders | 0.250 |
| | Mean of each subset | | | 7.3 | | | 7.7 | Variance of the subset means | 0.028 |

## 6.3 Numerical demonstration of Proposition 4

Table 6 illustrates the results of Proposition 4. We randomly generated 12 orders [8 7 9 5 10 10 10 5 9 7 5 9] and 12 demands [9 8 5 9 9 8 10 8 8 10 5 9]. The non-aggregated bullwhip ratio is calculated as 1.47. First, we divided the data into 6 subsets with 2 values in each subset ($M = 6, k = 2$). The bullwhip ratio calculated from the subset means is 1.46, which is very close to the within-subset bullwhip ratio of 1.48. Consequently, the aggregated bullwhip ratio of 1.46 is nearly identical to the non-aggregated bullwhip ratio, thus validating Formula (8). Next, we divided the order and demand datasets into 4 subsets, each containing 3 values ($M = 4, k = 3$). In this case, the bullwhip ratio based on the subset means is 0.82, which is smaller than the within-subset bullwhip ratio of 1.56. As a result, the aggregated bullwhip ratio of 0.82 is smaller than the non-aggregated bullwhip ratio, thereby validating Formula (7). Finally, we divided the datasets into 3 subsets with 4 values each ($M = 3, k = 4$). Here, the bullwhip ratio from the subset means is 2.38, which is greater than the within-subset bullwhip ratio of 1.40. Consequently, the aggregated bullwhip ratio of 2.38 is higher than the non-aggregated bullwhip ratio, confirming the validity of Formula (6).

Table 6. Relationship between bullwhip ratios (time aggregated vs. non-aggregated)

| | | 1 | 2 | 3 | 4 | 5 | 6 | 7 | 8 | 9 | 10 | 11 | 12 | | | | Bullwhip ratio |
|---|---|---|---|---|---|---|---|---|---|---|---|---|---|---|---|---|---|
| Order volume | | 8 | 7 | 9 | 5 | 10 | 10 | 10 | 5 | 9 | 7 | 5 | 9 | Variance of non-aggregated orders | 3.64 | R_non_agg | 1.47 |
| Demand volume | | 9 | 8 | 5 | 9 | 9 | 8 | 10 | 8 | 8 | 10 | 5 | 9 | Variance of non-aggregated demands | 2.47 | | |
| m=6, k=2 (m=6 subsets, each subset with k=2 values) | Subset variance of orders | 0.25 | | 4.00 | | - | | 6.25 | | 1.00 | | 4.00 | | Expectation of subset variances for orders | 2.58 | R_within (m=6, k=2) | 1.48 |
| | Subset variance of demands | 0.25 | | 4.00 | | 0.25 | | 1.00 | | 1.00 | | 4.00 | | Expectation of subset variances for demands | 1.75 | | |
| | Subset mean of orders | 7.50 | | 7.00 | | 10.00 | | 7.50 | | 8.00 | | 7.00 | | Variance of the subset means for orders | 1.06 | R_mean (m=6, k=2) | 1.46 |
| | Subset mean of demands | 8.50 | | 7.00 | | 8.50 | | 9.00 | | 9.00 | | 7.00 | | Variance of the subset means for demands | 0.72 | | |
| | Subset aggregation of orders | 15.00 | | 14.00 | | 20.00 | | 15.00 | | 16.00 | | 14.00 | | Variance of subset aggregation for orders | 4.22 | R_agg (m=6, k=2) | 1.46 |
| | Subset aggregation of demands | 17.00 | | 14.00 | | 17.00 | | 18.00 | | 18.00 | | 14.00 | | Variance of subset aggregation for demands | 2.89 | | |
| m=4, k=3 (m=4 subsets, each subset with k=3 values) | Subset variance of orders | | 0.67 | | | 5.56 | | | 4.67 | | | 2.67 | | Expectation of subset variances for orders | 3.39 | R_within (m=4, k=3) | 1.56 |
| | Subset variance of demands | | 2.89 | | | 0.22 | | | 0.89 | | | 4.67 | | Expectation of subset variances for demands | 2.17 | | |
| | Subset mean of orders | | 8.00 | | | 8.33 | | | 8.00 | | | 7.00 | | Variance of the subset means for orders | 0.25 | R_mean (m=4, k=3) | 0.82 |
| | Subset mean of demands | | 7.33 | | | 8.67 | | | 8.67 | | | 8.00 | | Variance of subset aggregation for demands | 0.31 | | |
| | Subset aggregation of orders | | 24.00 | | | 25.00 | | | 24.00 | | | 21.00 | | Variance of subset aggregation for orders | 2.25 | R_agg (m=4, k=3) | 0.82 |
| | Subset aggregation of demands | | 22.00 | | | 26.00 | | | 26.00 | | | 24.00 | | Variance of subset aggregation for demands | 2.75 | | |
| m=3, k=4 (m=3 subsets, each subset with k=4 values) | Subset variance of orders | | | 2.19 | | | | 4.69 | | | | 2.75 | | Expectation of subset variances for orders | 3.21 | R_within (m=3, k=4) | 1.40 |
| | Subset variance of demands | | | 2.69 | | | | 0.69 | | | | 3.50 | | Expectation of subset variances for demands | 2.29 | | |
| | Subset mean of orders | | | 7.25 | | | | 8.75 | | | | 7.50 | | Variance of the subset means for orders | 0.43 | R_mean (m=3, k=4) | 2.38 |
| | Subset mean of demands | | | 7.75 | | | | 8.75 | | | | 8.00 | | Variance of subset aggregation for demands | 0.18 | | |
| | Subset aggregation of orders | | | 29.00 | | | | 35.00 | | | | 30.00 | | Variance of subset aggregation for orders | 6.89 | R_agg (m=3, k=4) | 2.38 |
| | Subset aggregation of demands | | | 31.00 | | | | 35.00 | | | | 32.00 | | Variance of subset aggregation for demands | 2.89 | | |

## 7. Conclusions

### 7.1 Concluding remarks

Our paper provides a novel perspective and a systematic analysis on the measurement of the bullwhip ratio. We'll discuss findings of time aggregation, product aggregation, and seasonality respectively.

Findings on time aggregation. First, we use the law of total variance to decompose non-aggregated bullwhip ratio into four components, which provides a useful tool to analyze the effect of time aggregation on the bullwhip measurement. Second, the bullwhip ratio can be expressed by a weighted average of subset-wise bullwhip ratios based on Formula (4). Third, we interestingly find that Formula (3) contains components of the aggregated bullwhip ratio, which has been studied in Formula (5), offering a novel perspective to investigate the underlying factors of time aggregation. Fourth, time aggregation could increase, decrease or maintain the non-aggregated bullwhip ratio depending on difference context, correcting the masking effect of time aggregation studied by Chen and Lee (2012). Thus, we provide a rigorous underlying mechanism to interpret the conflicting results of time aggregation effect on bullwhip measurement in previous empirical studies (Lee et al., 2023; Chen and Lee, 2012; Yao et al., 2020; Cachon et al., 2007; Shan et al., 2013).

Findings on product aggregation. First, when aggregating uncorrelated products, the aggregated bullwhip ratio is a weighted average of bullwhip ratios from individual products, with corresponding demand variance as the weights, bounded by the minimum and maximum bullwhip ratios of the individual products. Second, when products can be correlated, the aggregated bullwhip ratio can be larger or smaller than the minimum and maximum bullwhip ratios of the individual products. Third, the principles of product aggregation extend beyond product level to broader organizational contexts, such as company-level and industry-level analyses. At the company level, the bullwhip ratio can be viewed as an aggregation of the ratios across different product lines.

Findings on seasonality. First, influence of seasonality on bullwhip ratio depends on the value of the de-seasonalized bullwhip ratio. When the de-seasonalized bullwhip ratio is larger than one, adding seasonality can decrease the bullwhip effect towards one. When the de-seasonalized bullwhip ratio is smaller than one, introducing seasonality could increase the bullwhip ratio. Second, seasonality can be viewed as a product with predictable demands, thus with a potential designed bullwhip ratio of one.

## 7.2 Managerial implications

Our findings have several managerial implications for industry practitioners.

First, when using time aggregation to calculate the bullwhip ratio, managers should be aware of that different time duration could lead to significantly varying results. Therefore, they should be very cautious when using the bullwhip ratio as a performance measure.

Second, when product aggregation is involved to calculate the bullwhip ratio, it is crucial for managers to distinguish the natural attributes of products, i.e., whether products are correlated or uncorrelated. If products are uncorrelated, applying a weighted average of bullwhip ratios will give managers a more precise bullwhip ratio. The aggregated bullwhip ratio of uncorrelated products will always fall between the minimum and maximum individual product ratios. Managers can use this knowledge to identify which products are driving the overall bullwhip effect and focus efforts on those with higher variability and greater influence on the aggregated bullwhip ratio.

Third, the principles of product aggregation can extend to higher organizational levels, such as company-wide or industry-wide analysis. High-level managers can use these principles to assess the bullwhip effect across multiple product lines or business units, allowing for a more comprehensive view of supply chain performance and enabling strategic adjustments at the macro level.

Fourth, seasonality can be viewed as a product with predictable demand patterns with a designed bullwhip ratio of one. Managers can capitalize on this predictability by implementing seasonality adjustments that align with their strategic goals, potentially smoothing out demand variability and improving overall supply chain stability.

Finally, managers could consider integrating time, product, and seasonality aggregation strategies to create a holistic approach to managing the bullwhip ratio. By understanding how these different types of aggregation interact and influence each other, managers can design more effective policies and practices that optimize supply chain performance across multiple dimensions.

If you have any questions or comments, please send them to me at:
**ma_mike@tongji.edu.cn**

**References**

Blitzstein, J.K. & Hwang, J. (2019). *Introduction to Probability: Texts in Statistical Science*. 2nd ed. Boca Raton, FL: CRC Press.

Bray, R.L. & Mendelson, H. (2012). Information transmission and the bullwhip effect: An empirical investigation. *Management Science*, 58(5), pp.860–875.

Cachon, G.P., Randall, T. & Schmidt, G.M. (2007). In search of the bullwhip effect. *Manufacturing & Service Operations Management*, 9(4), pp.457–479.

Chen, L. & Lee, H.L. (2009). Information sharing and order variability control under a generalized demand model. *Management Science*, 55(5), pp.781–797.

Chen, L. & Lee, H.L. (2012). Bullwhip effect measurement and its implications. *Operations Research*, 60(4), pp.771–784.

Cleveland, R.B., Cleveland, W.S., McRae, J.E. & Terpenning, I. (1990). STL: A seasonal-trend decomposition procedure based on loess. *Journal of Official Statistics*, 6(1), pp.3–73.

Dejonckheere, J., Disney, S.M., Lambrecht, M.R. & Towill, D.R. (2003). Measuring and avoiding the bullwhip effect: A control theoretic approach. *European Journal of Operational Research*, 147(3), pp.567–590.

Duan, Y., Yao, Y. & Huo, J. (2015). Bullwhip effect under substitute products. *Journal of Operations Management*, 36, pp.75–89.

Graves, S.C. (1999). A single-item inventory model for a nonstationary demand process. *Manufacturing & Service Operations Management*, 1(1), pp.50–61.

Ha, A.Y. & Tang, C.S. (2017). *Handbook of Information Exchange in Supply Chain Management*. Cham, Switzerland: Springer.

Horn, R.A. & Johnson, C.A. (1985). *Matrix Analysis*. Cambridge: Cambridge University Press, pp.176–180.

Jin, M., DeHoratius, N. & Schmidt, G. (2017). In search of intra-industry bullwhips. *International Journal of Production Economics*, 191, pp.51–65.

Lee, H.L., Padmanabhan, V. & Whang, S. (1997). Information distortion in a supply chain: The bullwhip effect. *Management Science*, 43(4), pp.546–558.

Lee, S., Park, S.J. & Sridhar, S. (2023). Variations of the bullwhip effect across foreign subsidiaries. *Manufacturing & Service Operations Management*, 25(1), pp.1–18.

Shan, J., Yang, S., Yang, S. & Zhang, J. (2013). An empirical study of the bullwhip effect in China. *Production and Operations Management*, 23(4), pp.537–551.

Strang, G. (2009). *Introduction to Linear Algebra*. 4th ed. Wellesley, MA: Wellesley-Cambridge Press.

Yao, Y., Duan, Y. & Huo, J. (2020). On empirically estimating bullwhip effects: Measurement, aggregation, and impact. *Journal of Operations Management*, 67(1), pp.5–30.